\documentclass[preprint]{aastex}  
\usepackage{amssymb}
\usepackage{natbib}
\newcommand {\hI} {\hbox{H\small I\normalsize}\,\,}
\newcommand {\ha} {H$\alpha$\,\,}  
\newcommand {\kms} {\,km\,s$^{-1}$\,}  
\newcommand {\M} {\mbox{${\mathcal M}$}}  
  
\newcommand {\msol} {\M$_\odot$\,}

\newcommand {\mlb} {(\M/L$_B$)$_\star$\,\,}

\newcommand {\mlsol}{\mbox{${\mathcal M}_\odot$/L$_{\odot}$}}

\renewcommand {\apj} {ApJ}

\renewcommand {\aj} {AJ}

\renewcommand {\mnras} {MNRAS}
\renewcommand {\aap} {A\&A}

\slugcomment{Submitted to {\it Astronomical Journal}}  
   
\begin{document}  
   
\title{ACCURATE DETERMINATION OF THE MASS DISTRIBUTION IN SPIRAL GALAXIES:\\
II. Testing the Shape of Dark Halos}

\author{S\'EBASTIEN BLAIS--OUELLETTE\altaffilmark{1}}
\affil{D\'epartement de physique and Observatoire du mont M\'egantic,  
Universit\'e de Montr\'eal, C.P. 6128, Succ. centre ville,  
Montr\'eal, Qu\'ebec, Canada. H3C 3J7 and\\  
Laboratoire d'Astrophysique de Marseille,  
2 Place Le Verrier, F--13248  Marseille Cedex 04, France\\
e--mail: blaisous@llnl.gov}
   
\author{PHILIPPE AMRAM}
\affil{Laboratoire d'Astrophysique de Marseille,  
2 Place Le Verrier, F--13248  Marseille Cedex 04, France\\  
e--mail: amram@observatoire.cnrs-mrs.fr} 

\author{CLAUDE CARIGNAN\altaffilmark{2}}
\affil{D\'epartement de physique and Observatoire du mont M\'egantic,  
Universit\'e de Montr\'eal, C.P. 6128, Succ. centre ville,  
Montr\'eal, Qu\'ebec, Canada. H3C 3J7\\  
e--mail: carignan@astro.umontreal.ca}   
  
\altaffiltext{1}{Present address: Inst. of Geophysics \& Planetary
Physics, Lawrence Livermore National Lab., CA 94550, USA}

\altaffiltext{2}{Visiting Astronomers, Canada--France--Hawaii Telescope,  
operated by the National Research Council of Canada, the Centre  
National de la Recherche Scientifique de France, and the  
University of Hawaii.}  

\begin{abstract}

New high resolution CFHT Fabry--Perot data, combined with published
VLA 21 cm observations are used to determine the mass distribution of
NGC 3109 and IC 2574. The multi-wavelength rotation curves allow to
test with confidence different dark halo functional forms from the
pseudo-isothermal sphere to some popular halo distributions motivated
by CDM N-body simulations. It appears that the density distributions
with high central concentration, predicted by these simulations, are
very hard to reconcile with rotation curves of late type
spirals. Modified Newtonian Dynamics (MOND) is also considered as a
potential solution to missing mass and tested the same way. Using
the higher resolution \ha data, and new \hI data for NGC 3109, one
can see that MOND can reproduce in details the rotation curves of IC
2574 and NGC 3109. However, the value for the MOND universal constant
is $\sim$2 times larger than the value found for more massive spirals.

\end{abstract}


\section{Introduction}  
\renewcommand{\thefootnote}{\alph{footnote}}
  
Over the last 30 years, rotation curves have been a
very efficient tool to study the mass distribution in spiral
galaxies. They clearly brought to light the important discrepancy
between the luminous mass and the gravitational mass that has led to
the supposition of a large amount of dark matter in the Universe. The
now commonly accepted picture is that this unseen matter takes the
form of large halos in the center of which galaxies are
embedded. Alternatively, the gravitational attraction could deviate
from the pure Newtonian force at very low acceleration so that no dark
matter is necessary.
  
The resolution reached by N-body simulations of the cosmic evolution
of dark halos \citep{NFW96,NFW97,Fukushige97,Moore98,Kravtsov98} allows
one to predict the inner part of halo density profiles. In theory,
these profiles could be directly compared to the ones deduced from
modeling the rotation curves. Unfortunately, the sensitivity of the
rotation curves to the exact density profile of the halos is quite low
and one must use the highest sensitivity and the highest resolution
possible to arrive at useful comparisons \citep[ hereafter paper
I]{paperI}.
  
This study investigates primarily, in the context of Newtonian
gravitation, the density profiles of the dark halos of two late type
spiral galaxies: NGC 3109 and IC 2574. As late type, their bulge is
minimal making them well suited objects for sensitive mass
distribution studies because of the reduced uncertainties due to the
negligible spheroid contribution in the inner parts. Mass models of
NGC 3109 based on radio observations have already been presented by
\cite{Jobin90} while the dark matter distribution in IC 2574 has been
studied by \citet{Martimbeau94}. New high resolution Fabry-Perot
observations of the \ha emission line are used in combination with
published 21 cm data to form accurate multi--wavelength rotation
curves. These curves are used to model the mass distribution in the
galaxies. The models include a stellar disk, a gaseous component and a
dark spherical halo.

Moreover, NGC 3109 has often been presented as a test for the
Modified Newtonian Dynamics \citep{Milgrom83a}. As noted in
\citet{BroeilsPhd}, the inner part of the rotation curve is crucial to
draw any conclusion on the MOND fit to this galaxy. Therefore, the
mass models of the two galaxies based on the MOND assumption are
presented as well.
  
The CFHT Fabry-Perot observations are described in Section
\ref{sec:fpdata3} while the mass models of the two galaxies, to which
we add NGC 5585 and NGC 3198 from paper I, are discussed in Section
\ref{sec:massmod}.

\section{New Fabry--Perot Observations}  
\label{sec:fpdata3}  

The Fabry-Perot observations of the \ha emission line were obtained in
February 1994 at the Canada--France--Hawaii Telescope (CFHT). The
etalon (CFHT1) was installed in the CFHT's Multi--Object Spectrograph
(MOS).  A narrow--band filter ($\Delta \lambda \approx$ 12\,\AA),
centered at $\lambda_0$ = 6565.5\,\AA\, for NGC 3109 (V$_{sys}
\approx$ 402 \kms) and at $\lambda_0$ = 6559.5\,\AA\, for IC 2574
(V$_{sys} \approx\, $53\kms) , was placed in front of the etalon.  The
available field with no vignetting was $\approx$ 8.5\arcmin $\times$
8.5\arcmin, with 0.314\arcsec\,px$^{-1}$. The pixel were binned 3 by 3 to
increase the signal-to-noise ratio and minimize the readout time. The
free spectral range of 5.66\,\AA\, (259 \kms) was scanned in 27 (plus one
overlapping) channels, giving a sampling of 0.21\,\AA\, (9.6 \kms) per
channel. Eight minutes integration were spent at each channel
position. Table~\ref{tab:fpobs3} lists the complete observing
parameters.
  
Following normal de--biasing and flat--fielding with standard IRAF
procedures, a robust 3-D cosmic--ray removal routine, that tracks
cosmic rays by spatial (pixel--to--pixel) and spectral
(frame--to--frame) analysis, was applied. Ghost reflections were then
removed using the technique described in Paper I.  With the ADHOC
software package\footnote{http://www-obs.cnrs-mrs.fr/adhoc/adhoc.html},
photometric variations were corrected using the mean night sky
(background + emission lines) to calculate the corrections to apply to
each frame.  Then, a calibration based on a neon lamp
($\lambda$6598.95 \AA) was used to fix the zero point of the spectrum
at each pixel.

\subsection{NGC 3109}  

Table~\ref{tab:opt3109} gives the optical parameters of NGC 3109. In
order to get sufficient signal--to--noise throughout the velocity
fields, two different Gaussian spatial smoothings ($\sigma$=3.5 and 5
pixels) were performed on the cube (all the channels). Velocity maps
were then obtained from the intensity weighted means of the H$\alpha$
peaks to determine the radial velocity for each pixel. A final variable
resolution velocity map (Fig. \ref{fig:vf+mono_3109}, top) was built
keeping higher resolution for regions with originally higher
signal-to-noise. The full resolution \ha image is also shown on the
lower part of Fig. \ref{fig:vf+mono_3109}.

The rotation curve was obtained from the velocity field using two
different methods. First, an estimate was made using the task ROCUR
\citep{BegemanPhd, Cote91} in the AIPS package, where annuli, tilted
with respect to the plane of the galaxy (ellipses in the plane of
sky), are fitted to the velocity field, minimizing the dispersion
inside each annulus. The dynamical center, systemic velocity, position
angle and inclination were estimated this way.  Secondly, the ADHOC
package was used to fine--tune these parameters by comparison of both
sides of the galaxy and examination of the residual velocity field
\cite[see][for the detailed method]{Amram92b}. The systemic velocity
was found to be 402 \kms, very close to the value of 404 \kms found
with the \hI observations of \cite{Jobin90}.  Since a warp is clearly
present even in the optical velocity field, the final rotation curve
was derived with ROCUR leaving the inclination and the position angle
free to vary. In this context, the slightly high value of the first
velocity point can be understood as a product of the poor
determination of the inclination in the center of the
galaxy. Table~\ref{tab:rc_3109_ha} gives the full rotation curve at
20\arcsec~resolution, while Fig. \ref{fig:rc_3109} illustrates the
rotation curve as well as the variation of inclination and position
angle. Following \cite{Jobin90}, the rotation curve was corrected for
asymmetric drift.

Currently, no convention on the way to represent the errors
on rotation curves exists in the literature. Error bars are often
simply given as the velocity dispersion in the ring used at each
radius. However, warm gas is known to be more sensitive to its
environment than cold gas. Turbulence, local density variations
(like spiral arms, bars, etc.) and winds from stars and supernovae of
the young stellar forming regions in which the ionized gas is found,
increase its dispersion. This can lead to the paradox that the fewer
points you have (as in long--slit observations) the lower is your
dispersion and the smaller are your error bars. As a more direct probe
of the uncertainties on the measured potential, the difference between
the two sides of the galaxy is instead often used. Some authors will
add the error due to uncertainties on inclination and/or position
angles.

To differentiate clearly the source of errors, the error bars shown
here indicate the error on the mean in each ring ($\sigma/\sqrt{N}$)
while the solutions for each side of the galaxy are represented by
lines (continuous for the receding side and dashed for the
approaching). Their difference is a good estimate of the asymmetry and
large scale non--circular motions.

The velocity field of NGC 3109 shows many bubble--like features
probably due to intense star formation especially close to the center
of the galaxy. The effect of \ha bubbles are normally of two kinds:
first, if the front and back sides of a bubble are equally intense and
the medium transparent, a broadened or even, depending on the
resolution, a doubled \ha line can be seen. The true gravitational
rotation can still be deduced by averaging the two peaks. However, in
many cases only the front side of an expanding bubble can be seen,
making it difficult to retrieve the true rotational velocities.

In the case of NGC 3109, analysis of the velocity field shows a small
deviation from circular motion in the inner part of the galaxy. The
velocity of the single innermost point of the rotation curve is
slightly affected.  Despite some internal peculiar velocities,
apparently well averaged out, a relatively good agreement can be seen
between the two sets of data. The \hI velocities (where no beam
smearing correction has been applied) were just slightly
underestimated in the inner part. This is not really surprising
considering that beam smearing effects should diminish as the slope of
the inner rotation curve diminishes for a given beam width. The
multi--wavelength rotation curve is thus composed of the \ha data
points up to 410 \arcsec\, and of \hI velocities for the rest.
  
\subsection{IC 2574}    

A similar reduction procedure was applied to IC 2574
(Table~\ref{tab:opt2574}) and the velocity field is shown juxtaposed
to the monochromatic image in Fig.  \ref{fig:vf+mono_2574}. It is
clear that IC 2574 is more disrupted than NGC 3109. With such a patchy
velocity field, a determination of a reliable rotation curve with an
iterative method based on velocity dispersion in annuli turned out to
be impossible. Direct calculation of the rotation velocity for each
pixel was thus done using fixed values for the position angle and the
inclination. The dynamical parameters were found by comparing the two
sides of the galaxy and by analyzing the residual field. The resulting
rotation curve is presented in Table~\ref{tab:rc_2574_ha} based on a
systemic velocity of 53 \kms, a position angle of 52\arcdeg and an
inclination of 75\arcdeg, almost identical to the parameters found by
\cite{Martimbeau94} from the \hI observations.

The \ha rotation curve (Fig.  \ref{fig:rc_2574}) follows more or less
the same kind of perturbations as those seen in the \hI curve. For
example, when one compares the two sides of the \hI curve in the
Fig. 8 of \cite{Martimbeau94}: (i) the effect of a giant bubble can
clearly be seen at both wavelengths around 240\arcsec\, on the
approaching side; (ii) the giant north--eastern OB association clearly
shows up between 7\arcmin\, and 10\arcmin\, of the receding side;
(iii) on the other hand, some probable non-circular motions are seen
around 100\arcsec\, in the \ha but do not show up in the \hI.

Since the comparision of \hI and \ha data shows no sign of beam
smearing, there would be no reason here to use the disrupted \ha curve
instead of the 21 cm data as the probe of the gravitational potential
of the galaxy.

\section{Mass Models and Parameters of the Mass Distribution}
\label{sec:massmod}  
  
\subsection{Comparison of different models using standard gravity}  

The method used in this paper to model the mass distribution is a
slight generalization of the one described in \citet{Car85a}.  The
luminosity profile, if possible in the near infrared to probe the mass
dominant stellar component, is transformed into a mass distribution
for the stellar disk, assuming a variable but radially constant
mass--to--light ratio (\mlb) \citep{Casertano83,Car85a}. For the
contribution of the gaseous component, the \hI radial profile is used,
scaled by 1.33 to account for He. The difference between the observed
rotation curve and the contribution to the curve from the luminous
(stars \& gas) component is thus the contribution of the dark
component, which can be represented by a dark spherical halo.  There
are therefore three free parameters, the \mlb for the disk, and two
parameters for the dark halo: the central density $\rho_0$ and the
core radius $r_c$.  A best fit routine minimizes the $\chi^2$ in the
three dimension parameter space.

Many studies approximate later type spirals such as NGC 3109 and IC
2574 as being totally made of dark matter, neglecting the stellar and
gaseous components. This is a good approximation down to a certain
radius. However, the innermost parts of the rotation curves are
crucial to test the shape of the density profiles. Best fit models
that include the contributions of gas and stars are therefore used to
avoid an overestimation of the dark halo contribution. 

Following \citet[ hereafter KKBP]{Kravtsov98} and \cite{Zhao96}, we use
an even broader family of density profiles for the halo:
  
\begin{equation}  
\rho(r) = \frac{\rho_0}{(c+(r/r_0)^\gamma) \left(1+(r/r_0)^\alpha\right)^{(\beta-\gamma)/\alpha}}  
\label{eqn:halo}
\end{equation}
where $\rho_0$ and $r_0$ are a characteristic density and radius respectively, and c
can force the presence of a flat density core. The parameters $\alpha$, $\beta$ and
$\gamma$ determine the shape of the density profile.

One can either fit the value of (c, $\alpha, \beta, \gamma$) to a
particular density profile or set them to a desired value: (1,
$\alpha\neq$ 0, 2, 2) for a pseudo--isothermal sphere
\citep{BegemanPhd}; (0, 1, 3, 1) for a NFW type halo \citep{NFW96};
(1, 2, 3, 1) for halos with flat density cores proposed by
\cite{Burkert95} or (0, 2, 3, 0.2) as proposed by KKBP. These four
density profiles are presented in Fig. \ref{fig:dens_profiles}.
Profiles with non-constant density cores ($lim_{r \to 0}(\rho) \neq
\rho_0$), are defined as cuspy. For these profiles, the inner slope is
given by $-\gamma$, the outer slope by $\beta$ while $\alpha$ controls
the sharpness of the turnover point.

NGC 3109 is particularly well suited for dynamical studies: First
because its luminous component is minimal \cite[B--band photometry
from][]{Kent87} so that the uncertainties related to the unknown
mass--to--light ratio of the disk do not have a significant
impact. Second, this galaxy is close enough to allow direct distance
estimation via multicolor observations of a large number of Cepheids
\citep{Musella97}. The adopted distance is 1.36 Mpc.

In Fig. \ref{fig:mm_3109_hiha} and \ref{fig:mm_2574_hi}, best--fit
models are given for the rotation curves of NGC 3109 and IC 2574,
respectively. The high resolution data (used up to
410 \arcsec\, for NGC 3109) remove some uncertainties pointed out by
\citet{Navarro97} in the differences between \hI curves from
different generations of radio-telescopes (single dish and aperture
synthesis). It confirms the great difficulty of reconciling a cuspy
profile with $\gamma \geq 1$ and the rotation curve of a late type
spiral like NGC 3109. Either constant density core profiles or mildly
cuspy profiles with $\gamma \ll 1$ can fit the data adequately. It is
very difficult to discriminate between a density distribution with a
flat core from one with a mild cusp because $r_0$ can often be
stretched to a point where the two types of profile match. One has to
go to very small radii ($\lesssim \gamma r_0$) to really probe any
incompatibility.
  
Cuspy profiles with $\gamma \geq 1$ seem to appear generically in cold
dark matter (CDM) N-body simulations but many models have been
suggested that deviate somewhat from the standard CDM assumptions and
avoid the creation of a steep central density cusp. As noted by KKBP,
\citet{Syer98} showed that $\gamma$ is sensitive to the past merger
rate and to the power spectrum of the initial density fluctuations on
the scale of a galactic halo. Therefore a less active merger history
or a steeper power spectrum should lead to $\gamma <$ 1 on galactic
scales.  Recently however \citep{Kravtsov00}, KKBP withdrew their
N-body simulation results. Thus, their density profiles used here
should not be considered as a product of CDM simulations.

Self-interacting dark matter \citep{Spergel00} has also been suggested
to suppress the formation of high density dark matter cusps but the
implied inverse dependency of the core radius on the mass of the
galaxy \citep{Dalcanton00} is not observed in our sample.

Alternatively, in order to explain the presence of a flat density core
in dwarf spirals for standard, scale free CDM models, \cite{Navarro96}
suggested that a violent starburst could eject the gas and consequently
flatten the inner dark matter distribution. Semi-analytical
calculations give this scenario the right order of magnitude for a
low mass galaxy given a sufficient feedback efficiency \citep{vdBosch00}.

Another way to reconcile the too steep rotation curves without
invoking a flat density CDM core is to add a second dark component,
composed of compact objects, to the non--baryonic cold dark matter
\citep{Burkert97}. Of course this adds some degrees of freedom when
fitting rotation curves and a better fit is in this case somewhat
meaningless, making this hypothesis hard to test.  The present
knowledge of the importance and distribution of MAssive Compact Halo
Objects (MACHOs) in our own Galaxy allow them to account for around 20\%
(up to 50\% at a 95\% confidence level) of the galactic dark matter
\citep{Alcock00}.  In any case, this is barely enough to solve the
cusp problem. Indeed, the MACHOs would need a very finely tuned
distribution.
   
NGC 5585 is also well fitted by all the profiles but the NFW
(Fig. \ref{fig:mm_5585_hiha}). The slight difference between the \mlb
found here (0.85) and the one found in paper I (0.80) is due to the
use of equation \ref{eqn:halo} instead of the integration of the
appropriate Lane-Emden equation for the isothermal sphere.

In the case of NGC 3198 (Fig. \ref{fig:mm_3198_hiha}), all the
profiles are compatible. The \ha curve has been used up to 65
\arcsec\, to rectify the overcorrection of the beam smearing in
\cite{BegemanPhd}. The only remark to be made is that all the profiles
are too smooth to account for the small variations in the outer
rotation curve, even though they are present in both the \hI and the B
profile. The detailed gravitational interplay between luminous and
dark matter should be taken into account. The detailed results of the
mass models can be found in Table \ref{tab:mm}.

\subsection{Modified Newtonian Dynamics}  
  
\citet{Milgrom83a} proposed that a modification of Newtonian
gravitation in the low acceleration limit could mimic a large amount
of dark matter in spiral galaxies. The MOdified Newtonian Dynamics
(MOND) is a truly falsifiable theory: it contains only one parameter,
$a_0$, that is supposed to be a universal constraint. A lot of work
has been done using rotation curves for which the quality of the data
gives the best opportunity to test the theory
\citep[e.g.][]{Begeman91, Sanders96, Sanders98, McGaugh98}.

As a reminder, here is the MOND quantitative prescription. 
For purely circular motion, one can equate the centripetal
 and gravitational acceleration. In the Newtonian regime,
we simply get
\begin{equation}  
\frac{V^2}{R}=\frac{G M_T}{R^2}=g_N  
\end{equation}  
  
MOND states that the true  
force is given by
\begin{equation}  
\mu(g/a_0){\bf g=g}_N 
\label{eq:mond_presc} 
\end{equation}  
where $\mu(g/a_0)$ is an interpolation function that has the right  
asymptotic behavior: $\mu(g/a_0 \gg 1)\rightarrow 1$ and $\mu(g/a_0 \ll  
1)\rightarrow g/a_0$. 

The exact form of $\mu(g/a_0)$ has no impact on the mass models of very late
type spirals like NGC 3109 and IC 2574 where the gravitational acceleration is
well below $a_0$ at all radii but
\begin{equation}
\mu(x)=\frac{x}{\sqrt{1+x^2}} 
\label{eq:mu}
\end{equation}
is generally assumed to be the interpolation function.

In the limit of low acceleration the gravitational acceleration is thus given  
by $g=\sqrt {a_0g_N}$ and  
\begin{equation}  
\frac{V^2}{R}=\sqrt{\frac{G M_T}{R^2}a_0}  
\end{equation}  
or  
\begin{equation}  
V^4=G M_T a_0  
\end{equation}  
which naturally explains the asymptotic flatness of rotation curves  
and the Tully-Fisher relation.   

A best--fit method, similar to the one used for dark halos has been
applied. The stellar luminosity profile and the HI density profile are
this time transformed into a mass distribution following equations
(\ref{eq:mond_presc}) and (\ref{eq:mu}). The free parameter of the fit
is as always the $(M/L)_\star$ ratio, $a_0$ is considered as a
universal constant, but, since no fundamental theory exists as of yet
to give its true value, 2 different values are used in the
fits. First, the value found by \cite{Begeman91} that best fits their
highly selective and high quality sample of luminous spirals ( $1.2
\times 10^{-13}$ \kms). The second value is the one that best fits
each present galaxy individually
(Table~\ref{tab:mond_par}). Fig. \ref{fig:mond} shows the best fit for
each galaxy and Table~\ref{tab:mond_par} indicates the parameters when
using the best fitted a$_0$.
 
IC 2574, NGC 3109 and NGC 3198 are fairly well fitted by the MOND law
if the ``universal constant'' a$_0$ is allowed to vary by more than a
factor 2. The quality of the fit is especially good in the case of IC
2574, where most of the features of the rotation curve are reproduced
in the fit. On the other side, the MOND prescription has more
difficulties to account for the mass distribution of NGC 5585 whatever
the value of a$_0$. Being for a large part in the Newtonian regime,
the case of NGC 5585 is somewhat less significant as it probes more
the interpolation function than the MOND theory itself. In contrast,
NGC 3109 belongs fully to the MOND regime. In this case, using new
Australia Telescope Compact Array \hI data improve significantly the
MOND fit compared to the one obtained with the old VLA data (the
$\chi^2$ going from 5 to 2), which were missing 1/3 of the flux.

\section{Summary and Discussion}

With the four galaxies studied so far, it is clear that while there is
a good agreement between Fabry-Perot and \hI data for the shallow
rotation curves of the late type galaxies NGC 3109 and IC 2574, beam
smearing plays an important role for the steeper rotation curves.
Beam smearing thus seems to depend on at least two factors: the inner
slope of the rotation curve and the sampling of the curve, often
expressed as the ratio of the Holmberg radius to the beam width
\citep{BosmaPhd}. Bosma suggested a ratio greater than 6 to have a
reliable \hI curve. This ratio is about 7 for NGC 5585 and 30 for NGC
3109 and the latter is still slightly affected by beam smearing.

Using multi-wavelength rotation curves to determine the density
profiles of dark matter halos, it appears clearly that cuspy profiles
with inner logarithmic slope $\gamma \geq$ 1 do not match the
observations. The best fits are achieved using either a slightly cuspy
profile with a very shallow inner slope or a profile with a flat
density core. If current simulations correctly describe CDM
evolution, an additional process is needed to destroy or forbid the
formation of a central cusp. However, none of the actually proposed
scenarios stand out as the most plausible either because they are not
easily testable or their predictions do not appear clearly in the
data.

The Modified Newtonian Dynamics prescription can fit the rotation
curves of IC 2574, NGC 3109 and NGC 3198 obtained from the new high
resolution data if a$_0$ is free to vary by a factor of 2. In this
case, most of the features of the rotation curve are even reproduced
with the right amplitude. Of course, since there is almost no stellar
disk, at least in the part fully in the MOND regime, scaling the \hI
would produce a similar result. Can $H_2$ be involved here?
\citep{Pfenniger94}. On the other hand, the variations of the stellar
and gaseous components also possibly just follow a non--smooth dark
matter distribution.

There is always a danger to let what should be a universal constant
like a$_0$ vary to fit the data. However, since no underlying theory
exists yet, it is interesting to note that for these late type spirals,
a$_0$ seems systematically higher than for the earlier types.

\section{Conclusion}

The present work leads to the following conclusions:

$\bullet$ Observations of the kinematics of the ionized hydrogen in
NGC 3109 and IC 2574 are in good agreement with the previous
kinematical studies of atomic hydrogen. This implies that beam
smearing was limited in the HI data, although NGC 3109 was slightly
affected.

\smallskip

$\bullet$ Overall, beam smearing can be important even for ``good
sampling'' depending on the inner slope of the rotation curve.

\smallskip

$\bullet$ The CDM models with a inner density slope $\gamma \geq 1$
are not compatible with the data on NGC 3109 and IC 2574. Flat
density core models like the pseudo-isothermal sphere or a model with
a shallow inner density slope are compatible with the four galaxies in
our sample.

\smallskip

$\bullet$ With the exception of NGC 5585, the MOND prescription can
fit the rotation curves of the three other galaxies in our sample if
the universal constant a$_0$ is allowed to vary by more than a factor
of 2. The a$_0$ values found for the late type galaxies are
systematically higher than what was found previously for more massive
spirals.
 
\smallskip

The present examples give a good idea of the impact of higher
resolution rotation curves. There are however large unexplored regions
in terms of galaxy mass, surface brightness and morphological
types. It is thus imperative to extend this sample to earlier type
galaxies covering a large range in surface brightness. This would
give, among other things, the opportunity to study precisely at what
point the rotation curves stop agreeing with N-body simulations.

\bigskip\bigskip

We would like to thank the staff of the CFHT for their support during
the data acquisition and Daniel Durand from the Hertzberg Institute
for Astrophysics who helped with data acquisition. We also warmly
thank Jacques Boulesteix for fruitful discussion on Fabry-Perot
reduction and Piotr Popowski for valuable comments. CC acknowledges
grants from NSERC (Canada) and FCAR (Qu\'ebec).

\newpage


\newpage



\figcaption[Blais-Ouellette.fig1.eps]{ Velocity field and \ha image of
NGC 3109. North is up and East is left. The X and the grey line
indicate the kinematic center and the axis of separation between the
approaching and receding sides. \label{fig:vf+mono_3109}}

\figcaption[Blais-Ouellette.fig2.eps]{ {\sl Top} \ha rotation curve of 
NGC 3109 (open circles) compared to the \hI rotation curve (filled
circles) from \cite{Jobin90}. The approaching and receding sides are
respectively represented by the dashed and continuous lines; {\sl
bottom} variation with radius of the position angle and inclination for the \ha
data. See the text for details on curves and error bars.
\label{fig:rc_3109}}



\figcaption[Blais-Ouellette.fig3.eps]{ Velocity field and \ha image of
IC 2574. North is up and East is left. The X and the grey line
indicate the kinematic center and the axis of separation between the
approaching and receding sides.
\label{fig:vf+mono_2574}}

\figcaption[Blais-Ouellette.fig4.eps]{ \ha rotation curve of IC 2574 (open circles) 
compared to the \hI rotation curve (filled circles) from
\cite{Martimbeau94}.  The approaching and receding sides of the \ha
curve are respectively represented by the dashed and continuous lines.
\label{fig:rc_2574}}


\figcaption[Blais-Ouellette.fig5.eps]{ Density profiles of the four models. {\sl solid}:  
pseudo-isothermal sphere, {\sl dot}: KKBP, {\sl long dash}: Burkert,
{\sl short dash}: NFW. 
\label{fig:dens_profiles}}



\figcaption[Blais-Ouellette.fig6.eps]{ Best fit mass models for NGC 3109 using 
the \ha rotation curve up to 2.7 kpc and the \hI rotation curve for
the rest. The dark halos density profiles are {\sl top--left}:
pseudo-isothermal sphere, {\sl top--right}: KKBP, {\sl bottom--left}:
Burkert, {\sl bottom--right}: NFW.
\label{fig:mm_3109_hiha}}

\figcaption[Blais-Ouellette.fig7.eps]{ Mass models of IC 2574 using the \hI rotation curve only. 
\label{fig:mm_2574_hi}}

\figcaption[Blais-Ouellette.fig8.eps]{ Mass models of NGC 5585 using the \ha rotation curve 
up to 3.5 kpc and \hI for the rest.
\label{fig:mm_5585_hiha}}

\figcaption[Blais-Ouellette.fig9.eps]{ Mass models of NGC 3198 using the \ha rotation curve 
up to 2.9 kpc and \hI for the rest.
\label{fig:mm_3198_hiha}}


\figcaption[Blais-Ouellette.fig10.eps]{ Best fit mass models using
MOND. Two different values of a$_0$ are used: the dotted line used
a$_0$= 1.2 $\times 10^{-13}$ km/s$^2$ from \cite{Begeman91}; the
continuous line used the best fit value (see
Table~\ref{tab:mond_par}). The filled circles represent the data used
in the fits (\hI+\ha for NGC 3109 and 5585 and \hI only for IC 2574
and NGC 3198). The open circles indicate the \ha velocities unused in
the fits.
\label{fig:mond}}


\begin{deluxetable}{lr}
\tablecaption{Parameters of the Fabry--Perot observations.\label{tab:fpobs3}}
\tablewidth{0pt}
\tablehead{}
\startdata
Dates of observations        &           February 21 and 22, 1994    	\\  
Telescope                   &                        3.6\,m CFHT  	\\  
Instrumentation: \hspace{8cm}& 						\\  
~~~~~Focal plane instrument &		MOSFP				\\  
~~~~~CCD detector           &            2048\,$\times$\,2048 Loral3,  
                                         $\sigma$ = 8\,e$^{-1}$   	\\  
~~~~~Fabry--Perot etalon    &           Scanning QW1162 (CFHT1) 	\\  
~~~~~Interference order     &            1155 @   
			$\lambda_{\scriptscriptstyle N\!E\!O\!N}$ 	\\  
~~~~~Mean Finesse in the field &  	 12				\\  
~~~~~Calibration lamp       &        Neon ($\lambda$ = 6598.95\,\AA)	\\  
Duration                    &						\\  
NGC 3109                    &						\\  
~~~~~Per channel	    &            7.28\,min/channel 		\\  
~~~~~Total                  &		 3\,h~24\,min			\\  
~~~~~Filter                 &            $\lambda_0$ = 6565.5\,\AA, 
					     $\Delta \lambda$ = 12\,\AA \\  
IC 2574                     &						\\  
~~~~~Per channel	    &            8.05\,min/channel 		\\  
~~~~~Total                  &		 3\,h~45\,min			\\  
~~~~~Filter                 &            $\lambda_0$ = 6559.5\,\AA, 
					     $\Delta \lambda$ = 12\,\AA \\  
Spatial Parameters:  	    &						\\  
~~~~~Field size             &            8.5$'$\,$\times$\,8.5$'$   	\\  
~~~~~Pixel scale            &            0.314$''$\,pix$^{-1}$      	\\  
Spectral Parameters: 	    &						\\   
~~~~~Number of channels     &                             27   		\\  
~~~~~Free spectral range    &     5.66\,\AA\ (259\,km\,s$^{-1}$) 	\\  
~~~~~Sampling    	    &     0.21\,\AA\ (9.6\,km\,s$^{-1}$)/channel \\
\enddata
\end{deluxetable}

\begin{deluxetable}{lr}
\tablecaption{Parameters of NGC 3109.\label{tab:opt3109}}  
\tablewidth{0pt}
\tablehead{}
\startdata
Morphological Type\tablenotemark{a}		&SBm			\\  
RA (J2000.0)			  &10$^{\rm h}$ 03$^{\rm m}$ 06\fs 6	\\  
Dec (J2000.0)		 		&-26\arcdeg 09\arcmin 32\arcsec  \\  
l						&262 \fdg 1		\\  
b						&23 \fdg 1		\\  
Adopted distance (Mpc)\tablenotemark{b}		&1.36 			\\  
						&(1\arcmin\ $\simeq 0.4$ kpc)\\  
Mean axis ratio, q = b/a\tablenotemark{c}		&0.28 $\pm 0.02$\\  
Inclination, i\tablenotemark{c}&		75 \arcdeg $\pm 2$ \arcdeg \\  
Isophotal major diameter, D$_{25}$\tablenotemark{c}	&14.4 \arcmin	\\  
Major axis PA\tablenotemark{c}			&93 \arcdeg $\pm 2$\arcdeg\\  
Exponential scale length (kpc)\tablenotemark{c} &1.2			\\  
Holmberg radius, R$_{\rm HO}$\tablenotemark{c}  & 13.3 \arcmin		\\  
Absolute magnitude, M$_B$\tablenotemark{c}	&--16.35 		\\  
Total luminosity, L$_B$			       &$5.2 \times 10^8$ L$_{\odot}$\\  
Helio. radial velocity (\kms)\tablenotemark{c}	&404 $\pm 3$		\\
\enddata
\tablenotetext{a}{\cite{RC3}}      
\tablenotetext{b}{\cite{Musella97}}
\tablenotetext{c}{\cite{Jobin90}}  
\end{deluxetable}  

\begin{deluxetable} {c c c c c} 
\tablecaption{Optical rotation curve of NGC 3109 at 20\arcsec\, binning 
from ROCUR. The two sides and the global rotational velocities are 
computed independently. The later are corrected for asymmetric drift.
\label{tab:rc_3109_ha}} 
\tabletypesize{\small}
\tablecolumns{5}
\tablewidth{0pt}
\tablehead{ 
\colhead{R$_{sides}$} & \colhead{V$_{app}$} & \colhead{V$_{rec}$} & \colhead{R} & \colhead{V} \\ 
\colhead{\arcsec} & \colhead{\kms} & \colhead{\kms} & \colhead{\arcsec} & \colhead{\kms} 
}
\startdata
35 & 8 $\pm$ 1 & 3 $\pm$ 1 & 30 & 10 $\pm$ 2 \\ 
65 & 6 $\pm$ 1 & & 70 & 10 $\pm$ 1 \\ 
95 & 15 $\pm$ 1 & & 90 & 15 $\pm$ 1 \\ 
& & & 110 & 15 $\pm$ 1 \\ 
125 & 14 $\pm$ 1 & & 130 & 15 $\pm$ 1 \\ 
155 & 22 $\pm$ 2 & & 150 & 20 $\pm$ 1 \\ 
& & & 170 & 25 $\pm$ 1 \\ 
185 & 25 $\pm$ 1 & & 190 & 25 $\pm$ 1 \\ 
215 & 26 $\pm$ 1 & & 210 & 26 $\pm$ 1 \\ 
& & & 230 & 32 $\pm$ 1 \\ 
245 & 32 $\pm$ 1 & & 250 & 33 $\pm$ 1 \\ 
275 & 34 $\pm$ 1 & 33 $\pm$ 1 & 270 & 37 $\pm$ 2 \\ 
& & & 290 & 35 $\pm$ 2 \\ 
305 & 35 $\pm$ 1 & 32 $\pm$ 135 & 310 & 37 $\pm$ 2 \\ 
335 & 39 $\pm$ 2 & 37 $\pm$ 1 & 330 & 39 $\pm$ 2 \\ 
& & & 350 & 41 $\pm$ 4 \\ 
365 & 46 $\pm$ 6 & 40 $\pm$ 8 & 370 & 49 $\pm$ 10 \\ 
395 & 52 $\pm$ 9 & 42 $\pm$ 4 & 390 & 47 $\pm$ 2 \\ 
& & & 410 & 48 $\pm$ 1 \\ 
425 & 46 $\pm$ 2 & 45 $\pm$ 11 & \\ 
\enddata
\tablecomments{derived with V$_{sys}$ = 402 \kms (see text)} 
\end{deluxetable}

\begin{deluxetable}{lr}
\tablecaption{Optical parameters of IC 2574.\label{tab:opt2574}}  
\tablewidth{0pt}
\tablehead{}
\startdata
Morphological Type\tablenotemark{a}		&SABm			\\  
RA (J2000.0)			  &10$^{\rm h}$ 28$^{\rm m}$ 21\fs 2	\\  
Dec (J2000.0)		 		&68\arcdeg 24\arcmin 43\arcsec  \\  
l						&140 \fdg 2		\\  
b						&43 \fdg 6		\\  
Adopted distance (Mpc)\tablenotemark{b}		&3.0 			\\  
						&(1\arcmin\ $\simeq 0.8$ kpc)\\  
Mean axis ratio, q = b/a\tablenotemark{b}		&0.48 $\pm 0.06$\\  
Inclination, i\tablenotemark{b}&		75 \arcdeg $\pm 3$ \arcdeg \\  
Isophotal major diameter, D$_{25}$\tablenotemark{b}	&9.8 \arcmin	\\  
Major axis PA\tablenotemark{b}			&52 \arcdeg $\pm 6$\arcdeg\\  
Exponential scale length (kpc)\tablenotemark{b} & 2.2			\\  
Holmberg radius, R$_{\rm HO}$\tablenotemark{b}  & 8.6 \arcmin		\\  
Absolute magnitude, M$_B$\tablenotemark{b}	&--16.77 		\\  
Total luminosity, L$_B$			       &$8.0 \times 10^8$ L$_{\odot}$\\  
Helio. radial velocity (\kms)\tablenotemark{b}	&58 $\pm 3$		\\
\enddata
\tablenotetext{a}{\cite{RC3}}         
\tablenotetext{b}{\cite{Martimbeau94}}
\end{deluxetable} 

\begin{deluxetable}{c c c c c c c c}
\tabletypesize{\small}
\tablecolumns{8}
\tablewidth{0pt}
\tablecaption{Optical rotation curve of IC 2574 at 9.4\arcsec\, binning from ADHOC. 
The two sides are computed independently and the total is their mean 
averaged by the number of points. The velocity dispersion and the 
number of points appear for both sides while the error is the  
$\sigma/\sqrt{N}$ of the sides, added in quadrature. \label{tab:rc_2574_ha}}
\tablehead{ 
\colhead{R} & \colhead{N$_{app}$} & \colhead{V$_{app}$} & \colhead{$\sigma_{ring}$} 
& \colhead{N$_{rec}$} & \colhead{V$_{rec}$} & \colhead{$\sigma_{ring}$} & \colhead{V} \\ 
\colhead{arcsec} && \colhead{\kms} & \colhead{\kms} & & \colhead{\kms} 
& \colhead{\kms} & \colhead{\kms} 
}
\startdata
8 & & & & 2 & 22 & 8 & 22 $\pm$ 5 \\ 
12 & & & & 3 & 34 & 9 & 34 $\pm$ 5 \\ 
15 & 22 & 8 & 20 & & & & 8 $\pm$ 4 \\ 
24 & 36 & 15 & 10 & 79 & 10 & 7 & 11 $\pm$ 1 \\ 
33 & 26 & 12 & 12 & 81 & 3 & 27 & 5 $\pm$ 3 \\ 
43 & 22 & 2 & 10 & & & & 2 $\pm$ 2 \\ 
52 & 62 & 10 & 15 & 144 & 3 & 22 & 5 $\pm$ 2 \\ 
61 & 101 & 17 & 22 & 206 & 6 & 17 & 10 $\pm$ 2 \\ 
70 & 139 & 27 & 24 & 228 & 9 & 16 & 16 $\pm$ 1 \\ 
80 & 151 & 30 & 29 & 200 & 14 & 14 & 21 $\pm$ 2 \\ 
89 & 113 & 42 & 37 & 202 & 16 & 16 & 25 $\pm$ 2 \\ 
99 & 62 & 43 & 23 & 174 & 29 & 13 & 33 $\pm$ 2 \\ 
108 & 75 & 33 & 23 & 157 & 30 & 16 & 31 $\pm$ 2 \\ 
118 & 78 & 31 & 25 & 144 & 32 & 20 & 32 $\pm$ 2 \\ 
127 & 72 & 33 & 27 & 137 & 31 & 19 & 32 $\pm$ 2 \\ 
137 & 53 & 53 & 28 & 206 & 25 & 12 & 30 $\pm$ 2 \\ 
146 & 40 & 66 & 26 & 423 & 26 & 9 & 30 $\pm$ 1 \\ 
155 & 41 & 79 & 33 & 646 & 28 & 8 & 31 $\pm$ 1 \\ 
165 & 45 & 54 & 40 & 739 & 31 & 13 & 32 $\pm$ 2 \\ 
174 & 68 & 37 & 46 & 734 & 32 & 11 & 33 $\pm$ 2 \\ 
183 & 81 & 25 & 18 & 712 & 34 & 10 & 33 $\pm$ 1 \\ 
193 & 94 & 30 & 15 & 593 & 36 & 6 & 35 $\pm$ 1 \\ 
202 & 102 & 31 & 22 & 510 & 36 & 8 & 35 $\pm$ 1 \\ 
212 & 141 & 32 & 29 & 457 & 37 & 12 & 36 $\pm$ 1 \\ 
221 & 193 & 34 & 24 & 410 & 40 & 13 & 38 $\pm$ 1 \\ 
230 & 155 & 30 & 22 & 309 & 44 & 16 & 40 $\pm$ 1 \\ 
240 & 147 & 42 & 34 & 418 & 48 & 12 & 47 $\pm$ 2 \\ 
249 & 139 & 42 & 34 & 487 & 48 & 14 & 46 $\pm$ 1 \\ 
259 & 185 & 43 & 28 & 497 & 46 & 15 & 45 $\pm$ 1 \\ 
268 & 190 & 40 & 21 & 372 & 43 & 12 & 42 $\pm$ 1 \\ 
278 & 138 & 44 & 23 & 234 & 40 & 21 & 41 $\pm$ 2 \\ 
287 & 245 & 39 & 20 & 262 & 34 & 19 & 36 $\pm$ 1 \\ 
296 & 287 & 44 & 13 & 258 & 34 & 28 & 39 $\pm$ 1 \\ 
306 & 289 & 42 & 10 & 279 & 45 & 24 & 44 $\pm$ 1 \\ 
314 & 279 & 38 & 11 & 218 & 42 & 23 & 40 $\pm$ 1 \\ 
325 & 188 & 36 & 20 & 172 & 40 & 24 & 38 $\pm$ 2 \\ 
334 & 347 & 37 & 20 & 195 & 45 & 31 & 40 $\pm$ 2 \\ 
343 & 258 & 36 & 20 & 177 & 45 & 29 & 40 $\pm$ 2 \\ 
352 & 159 & 40 & 13 & 183 & 47 & 27 & 44 $\pm$ 2 \\ 
362 & 182 & 42 & 8 & 181 & 44 & 24 & 43 $\pm$ 1 \\ 
372 & 327 & 44 & 15 & 175 & 46 & 23 & 45 $\pm$ 1 \\ 
381 & 440 & 43 & 12 & 205 & 47 & 25 & 45 $\pm$ 1 \\ 
390 & 255 & 40 & 18 & 217 & 53 & 28 & 46 $\pm$ 2 \\ 
400 & 225 & 39 & 20 & 223 & 49 & 25 & 44 $\pm$ 2 \\ 
409 & 162 & 41 & 29 & 291 & 52 & 26 & 48 $\pm$ 2 \\ 
418 & 174 & 47 & 20 & 360 & 57 & 29 & 54 $\pm$ 1 \\ 
428 & 162 & 50 & 42 & 424 & 60 & 32 & 57 $\pm$ 2 \\ 
437 & 159 & 52 & 33 & 446 & 58 & 36 & 57 $\pm$ 2 \\ 
446 & 171 & 46 & 31 & 367 & 55 & 27 & 52 $\pm$ 2 \\ 
456 & 167 & 52 & 25 & 317 & 45 & 24 & 48 $\pm$ 2 \\ 
466 & 173 & 55 & 23 & 343 & 48 & 28 & 51 $\pm$ 2 \\ 
475 & 224 & 55 & 29 & 332 & 52 & 30 & 53 $\pm$ 2 \\ 
484 & 240 & 59 & 45 & 339 & 54 & 28 & 56 $\pm$ 2 \\ 
494 & 214 & 60 & 28 & 363 & 59 & 28 & 59 $\pm$ 2 \\ 
503 & 171 & 64 & 36 & 434 & 60 & 22 & 61 $\pm$ 2 \\ 
512 & 126 & 77 & 33 & 437 & 58 & 31 & 62 $\pm$ 2 \\ 
522 & 96 & 70 & 15 & 504 & 59 & 37 & 61 $\pm$ 2 \\ 
531 & 87 & 63 & 12 & 519 & 63 & 36 & 63 $\pm$ 2 \\ 
540 & 85 & 73 & 32 & 511 & 62 & 37 & 64 $\pm$ 2 \\ 
550 & 90 & 81 & 45 & 537 & 64 & 41 & 66 $\pm$ 2 \\ 
559 & 65 & 71 & 30 & 505 & 65 & 39 & 66 $\pm$ 2 \\ 
568 & 36 & 77 & 17 & 549 & 68 & 36 & 68 $\pm$ 2 \\ 
578 & 17 & 85 & 25 & 553 & 64 & 35 & 65 $\pm$ 2 \\ 
587 & 18 & 81 & 4 & 517 & 63 & 35 & 64 $\pm$ 2 \\ 
595 & 3 & 79 & 4 & 433 & 55 & 42 & 55 $\pm$ 2 \\ 
606 & & & & 465 & 60 & 38 & 60 $\pm$ 2 \\ 
616 & & & & 496 & 67 & 36 & 67 $\pm$ 2 \\ 
625 & & & & 521 & 76 & 37 & 76 $\pm$ 2 \\ 
634 & & & & 531 & 79 & 35 & 79 $\pm$ 2 \\ 
645 & 8 & 60 & 78 & 518 & 82 & 36 & 82 $\pm$ 4 \\ 
653 & 19 & 101 & 75 & 513 & 82 & 41 & 82 $\pm$ 4 \\ 
663 & & & & 478 & 71 & 43 & 71 $\pm$ 2 \\ 
672 & & & & 476 & 70 & 36 & 70 $\pm$ 2 \\ 
682 & & & & 462 & 69 & 42 & 69 $\pm$ 2 \\ 
691 & & & & 465 & 78 & 38 & 78 $\pm$ 2 \\ 
700 & & & & 465 & 88 & 38 & 88 $\pm$ 2 \\ 
710 & & & & 460 & 87 & 39 & 87 $\pm$ 2 \\ 
719 & & & & 421 & 83 & 31 & 83 $\pm$ 1 \\ 
728 & & & & 348 & 85 & 29 & 85 $\pm$ 2 \\ 
738 & & & & 342 & 94 & 20 & 94 $\pm$ 1 \\ 
747 & & & & 323 & 94 & 20 & 94 $\pm$ 1 \\ 
757 & & & & 282 & 95 & 24 & 95 $\pm$ 1 \\ 
766 & & & & 280 & 93 & 25 & 93 $\pm$ 2 \\ 
776 & & & & 255 & 92 & 34 & 92 $\pm$ 2 \\ 
785 & & & & 268 & 97 & 33 & 97 $\pm$ 2 \\ 
794 & & & & 277 & 97 & 40 & 97 $\pm$ 2 \\ 
803 & & & & 215 & 99 & 46 & 99 $\pm$ 3 \\ 
813 & & & & 163 & 101 & 42 & 101 $\pm$ 3 \\ 
823 & & & & 115 & 110 & 32 & 110 $\pm$ 3 \\ 
869 & & & & 61 & 120 & 14 & 119 $\pm$ 2 \\ 
877 & & & & 24 & 103 & 12 & 102 $\pm$ 2 \\ 
\enddata
\tablecomments{derived with V$_{sys}$ = 53 \kms, i = 75 \arcdeg, PA = 52\arcdeg } 
\end{deluxetable}

\begin{deluxetable}{l l l c c c c c}
\tablecolumns{7}
\tablewidth{0pt}
\tablecaption{Parameters of the mass models.\label{tab:mm}}
\tablehead{\colhead{Model} & \colhead{Galaxy} & \colhead{Type} & \colhead{\mlb} 
& \colhead{r$_0$} & \colhead{$\rho_0$}& \colhead{c} &\colhead{$\chi^2$} \\
 & & & & \colhead{kpc} & \colhead{10$^{-3}$\msol/pc$^{-3}$} &
}
\startdata
\rule[0mm]{0mm}{4mm}ISO & IC 2574  & SABm  & 0.34 & 5.4  & 7.0 & 10 & 2.7 \\
    & NGC 3109 & SBm   & 0.04 & 2.4 & 2.5 & 6.6 & 2.8 \\
    & NGC 5585 & SABd  & 0.85 & 2.2 & 4.3 & 8.2 & 4.4 \\
    & NGC 3198 & SBc   & 4.8 & 2.5  & 5.7 & 9.2 &7.1\\
\rule[0mm]{0mm}{4mm}Burkert & IC 2574  &  & 0.6 & 8.0 & 8.4 & 11 & 2.5 \\
	& NGC 3109 &    & 0.35 & 4.1 & 2.5 & 6.6 & 2.9 \\
	& NGC 5585 &    & 0.086 & 4.0 & 4.0 & 8.0 & 8.2 \\
	& NGC 3198 &    & 5.0 & 5.8 & 7.4 & 10 & 7.7 \\
\rule[0mm]{0mm}{4mm}KKBP & IC 2574  &   & 0.25 & 9.2 & 5.0 & 8.8 & 2.4 \\
     & NGC 3109 &   & 0.40  & 4.5 & 1.5 & 5.2 & 2.9 \\
     & NGC 5585 &   & 0.86 & 3.9 & 5.6 & 9.2 & 7.4 \\
     & NGC 3198 &   & 6.0   & 9.2  & 1.2 & 4.7 & 9.1\\
\rule[0mm]{0mm}{4mm}NFW & IC 2574  & & 0.0 & 35 & 0.05 & 1.0 & 44 \\
    & NGC 3109 &   & 0.0 & 109 & 0.024 & 0.61 & 10 \\
    & NGC 5585 &   & 0.0 & 10.1 & 0.77 & 3.9 & 7.9 \\
    & NGC 3198 &   & 3.0 & 11.2 & 1.3 & 4.9 & 44 \\ 
\enddata
\tablecomments{Models parameters, r$_0$ and $\rho_0$, are respectively 
the caracteristic radius and caracteristic density. The concentration $c$ is 
calculated following \cite{NFW96} and is only exact in the NFW case. Because 
of the model dependance, c is only approximate for the other cases.} 
\end{deluxetable}

\begin{deluxetable}{l c c c c}
\tablecolumns{5}
\tablewidth{0pt}
\tablecaption{MOND parameters when using the best fitted a$_0$. \label{tab:mond_par}}
\tablehead{
\colhead{Galaxy} & \colhead{Distance} & \colhead{a$_0$} & \colhead{\mlb} & \colhead{$\chi^2$}\\
& \colhead{Mpc} & \colhead{$\times 10^{-13}$ km/s$^2$} && \colhead{\mlsol} \\ 
}
\startdata
IC 2574  & 3.0  & 2.0  & 0.02 & 1.5 \\
NGC 3109 & 1.36 & 2.7 & 0.12 & 5.0 \\
NGC 5585 & 6.2  & 2.5 & 0.34 & 5.9\\
NGC 3198 & 9.36 & 0.9  & 6.7 & 1.1\\ 
\enddata
\end{deluxetable}  

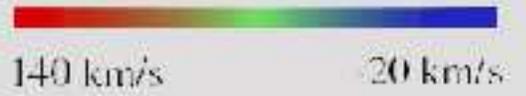
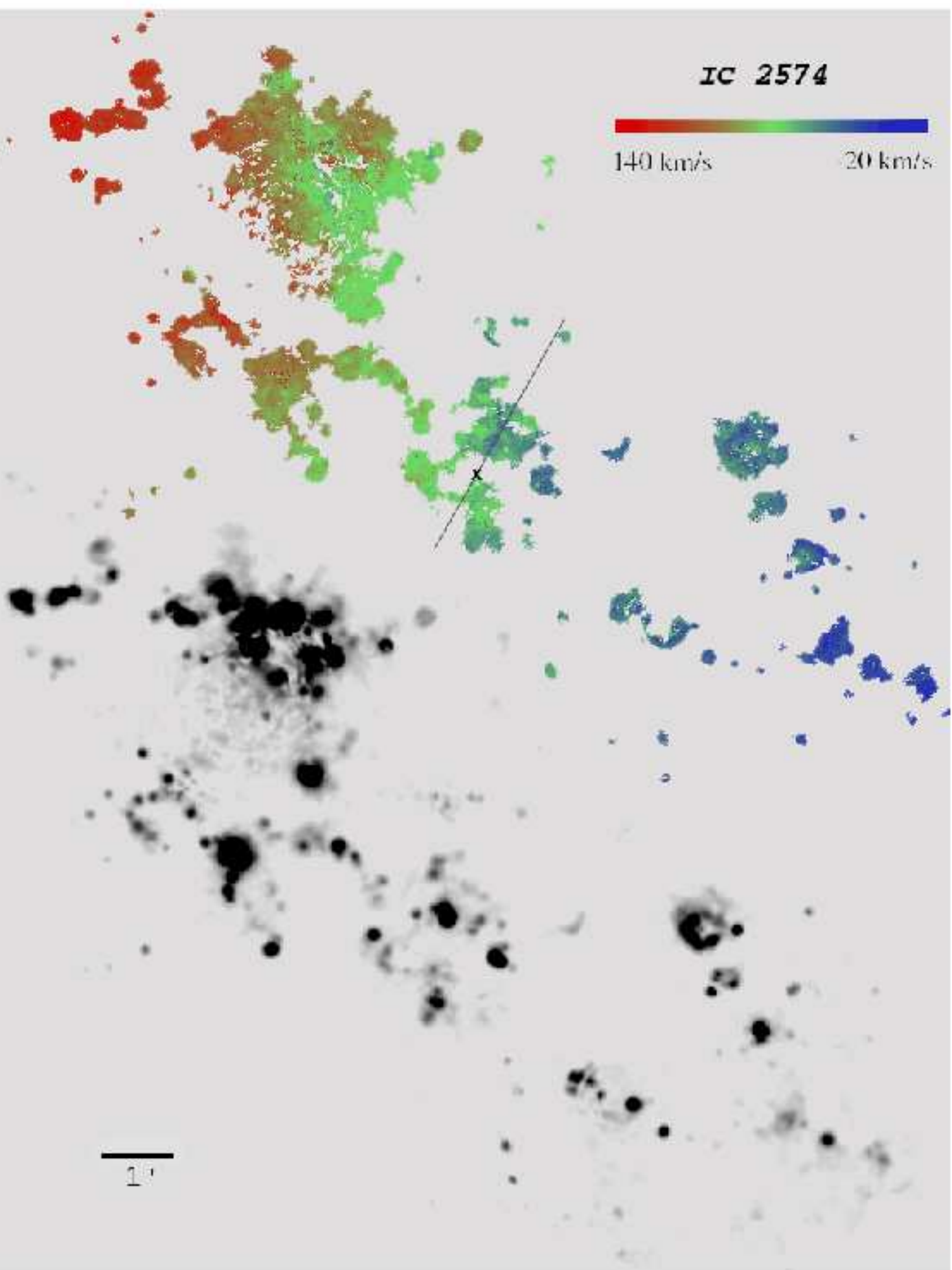
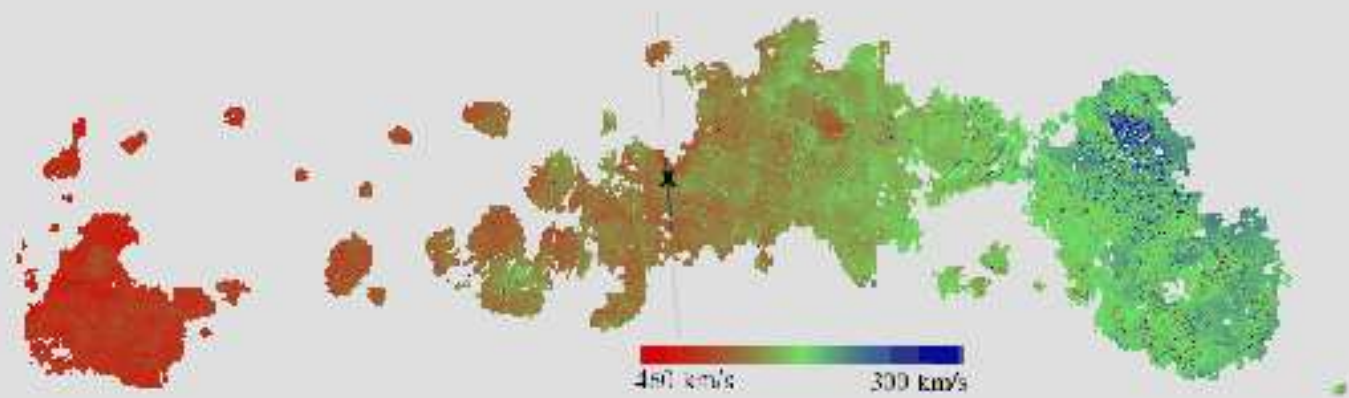
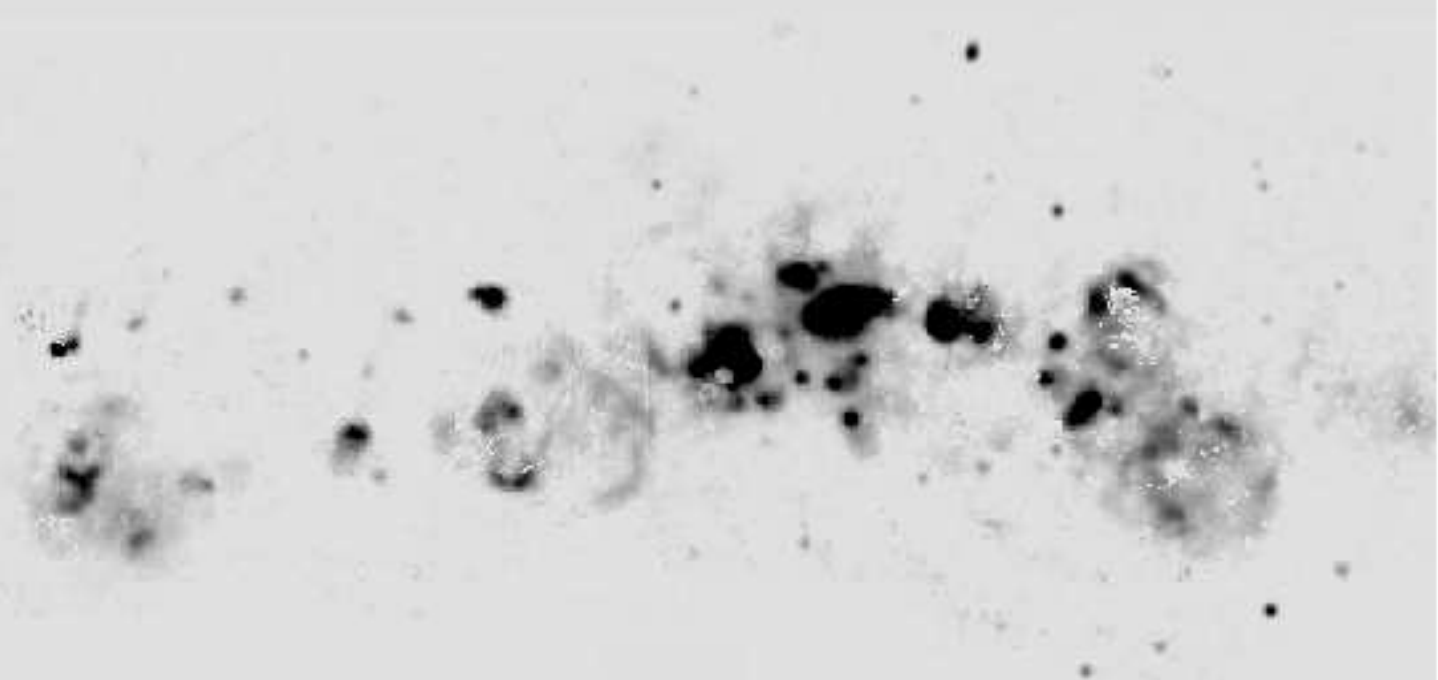
\end{document}